# Towards the Design of Effective Freehand Gestural Interaction for Interactive TV

Gang Ren[a,*], Wenbin Li[b] and Eamonn O'Neill[c]
[a]*School of Digital Arts, Xiamen University of Technology, No. 600 Ligong Road, 361021, Xiamen, China.*
[b]*Department of Computer Science, University College London, WC1E 6BT, London, UK*
[c]*Deparment of Computer Science, University of Bath, BA2 7AY, Bath, UK.*

**Abstract.** As interactive devices become pervasive, people are beginning to looking for more advanced interaction with televisions in the living room. Interactive television has the potential to offer a very engaging experience. But most common user tasks are still challenging with such systems, such as menu selection or text input. And little work has been done on understanding and supporting the effective design of freehand interaction with an TV in the living room. In this paper, we perform two studies investigating freehand gestural interaction with a consumer level sensor, which is suitable for TV scenarios. In the first study, we investigate a range of design factors for tiled layout menu selection, including wearable feedback, push gesture depth, target size and position in motor space. The results show that tactile and audio feedback have no significant effect on performance and preference, and these results inform potential designs for high selection performance. In the second study, we investigate a common TV user task of text input using freehand gesture. We design and evaluate two virtual keyboard layouts and three freehand selection methods. Results show that ease of use and error tolerance can be both achieved using a text entry method utilizing a dual circle layout and an expanding target selection technique. Finally, we propose design guidelines for effective, usable and comfortable freehand gestural interaction for interactive TV based on the findings.

## 1. Introduction

The increasing use of interactive TV in the living room brings novel opportunities and requirements for rich and engaging interactive experiences. The interactive TV now can deliver various interactive services such as on-demand delivery of video content from popular websites, gaming, shopping and even serve as the control center of a smart home. There are many different ways to interact with an interactive TV. The most common input method is using a traditional remote control, however, the remote normally offers only a limited set of buttons and does not lend itself to offering richer means of interaction. Other input methods used for computers or mobile devices can also be used with interactive TVs, such as keyboard, mouse and touch-sensitive displays. However, these input devices are usually intended to be installed close to or even contiguous with the screen so they are not suitable for many typical use scenarios with an interactive TV, where the user is often at a distance from the screen. Mobile devices such as phones or tablets can also be used to interact with remote displays, but configuration is often needed to connect the personal devices and this may not be convenient in some scenarios.

On the other hand, gestural input is increasingly popular, using hands-on input devices (e.g. Wii Remote) or freehand motion tracking by a camera (e.g. Microsoft Kinect). As gestural input moves beyond the home gaming settings in which it is already very popular, freehand gestural interaction, which has no need for hands-on input devices and so enables easier and more convenient "walk up and use" [1], is likely to become more important in interactions with TV in everyday settings.

Currently, however, it is still difficult to perform some common tasks such as option selection or text entry with freehand gestural interaction for interactive TV. For example, when a person is trying to select a video from many options, or search for a program or a video clip on an interactive TV, her interaction task may be challenging due to several factors including, for example, the relatively low resolution of many remote gesture sensors and the distance to the TV screen. We are therefore motivated to investigate effective freehand gestural interaction design for interactive TV.

*Corresponding author. Email: rengang@xmut.edu.cn

We first present a review of background and related work. We then report two studies focusing on 2 common interaction tasks using interactive TV, selection and text input, investigating the effects of a range of factors that may influence the design and usability of such interaction, including wearable feedback, push gesture depth, target size and position in motor space, as well as keyboard layout and selection techniques for text input. Drawing on our findings, we provide guidelines for freehand gestural interaction designers.

## 2. Background and Related Work

There is a quite long history of gestural interaction investigation and many various gesture input methods have been designed and investigated. Efforts were also made to summarize and classify gesture types [5, 6, 7]. For example, gesture styles can be classified as deictic, manipulation, gesticulation, semaphores and sign language [5]. Deictic and manipulative gestures are similar to pointing and manipulating used in physical word, thus no extra learning or training is required to use such gestures. Gesticulation accompanies speech and it needs no special learning either. On the other hand, a dictionary and even grammatical structures are required if using semaphoric gesture and sign language, so in a computer system, extra learning is needed for using these gesture types. Some criticism is made on semaphoric and sign language gesture styles. For example, such gestures are a step back to command-line interfaces [12] and natural user interfaces based on those gestures are not really natural [13]. Some arguments also indicated that gestural interaction should be based on well-designed metaphors rather than gesture design [14].

*2.1. The selection task*

Selection task is one of the most basic interaction task with a computer system. Various input devices can be used for selection tasks, including gesture inputs. Bobeth et al. [8] tested freehand menu selection for interactive TV with 4 different designs, and found that freehand gestures could be an appropriate way for older adults to control a TV. A selection task was also investigated in [9], where participants preferred freehand gestural pointing to using a handheld pointing device. Drawing different shapes in the air can also be used to select objects or menu items [10] with interactive TV, however, certain shapes are not easy to perform and remember, and have low recognition rates. User defined gestures for TV were also evaluated in [11]. The results showed that a pointing action was frequently used and a desktop interaction style, such as a push in mid-air to simulate clicking, was observed in many cases.

With the popularity of touch sensitive devices, people are very familiar with tapping on mobile phones to select from tile layout interfaces. Such "walk-up-and-use" experience could also be used for larger displays with freehand gestural interaction [1]. However, compared to other input methods such as a mouse or touch surface, freehand interaction typically lacks feedback, e.g. audio or tactile, which may reduce its effectiveness [2]. Research has suggested that freehand gesture interaction with menus could benefit from spatial auditory feedback via a headset [3], however, [4] showed that tactile feedback via a data glove had no benefit for selection tasks with pinch gestures.

Other design factors in freehand gestural selection, such as target size, have also been evaluated. For example, Vogel and Balakrishnan [2] tested freehand selection gestures with different target distances and target width. However, studies have mainly focused on sparse target layouts rather than tile layouts where objects are proximate to each other, and there has been little research on freehand gestural interaction with tile layout interfaces, and the available guidelines for the design of such interfaces are very limited.

*2.2. Text input task*

Internet connections bring large amount of online video content to TVs, and raise the requirements of searching those content quickly. Usually, the search is initialized by type keywords. Although at the moment the TV remote controller is still the most popular input device, but may not be the best option for text input, as indicated by [18]. They have compared a physical Qwerty keyboard to a remote con-trol for text entry task, and suggested that the performance of Qwerty keyboard is better. Still, many text input methods designed for TV remotes have been proposed and investigated [15, 16, 17].

Gestural text input methods have gained more attentions recently and there are many gestural text entry methods proposals and investigations lately. Most of these methods could also be used with interactive TV. For example, accelerometer has been used in gesture enabled virtual keyboard entry with user performance of 3.7 words per minute (wpm) in first time use and 5.4 wpm after 4 days' practice. Data gloves and fiducial markers are also popular input devices for stroke-based text entry methods, a study [20] shoed that users can reached 6.5 wpm without word completion after 2 weeks' practice. Text entry was also evaluated with large displays [21], and the results showed that Qwerty layout had the best performance (18.9 wpm), but the typing speed decreased significantly with more errors as the user moved away from the display.

Shape writing is another popular topic for gestural input investigations. With a stylus keyboard, Kristensson and Zhai [22] saw high performance in informal trials. However, another text entry test on mobile touch screens showed Qwerty was faster and more accurate than handwriting and shape writing text entry [23]. There are also some other text entry methods originally designed for stylus and touch screen can be redesigned for freehand text entry, such as FlowMenu [24] and Quikwriting [25, 26].

It is also possible to track freehand gesture with 3D cameras rather than holding a device in hands or wearing a data glove. Popular low-costs sensors include Microsoft Kinect[2] and ASUS Xtion[3]. Without asking users to hold devices in hands or attach markers, remote cameras have the advantage of enabling hand tracking in 3D world. However, such tracking method also suffers low resolution and tracking accuracy as only one low-cost camera is deployed: the resolution of the Microsoft Kinect Version 1 3D sensor is 3mm in the X-Y dimensions and 1cm in Z dimension from 2 meters away [27, 28].

Using freehand gesture for single character recognition can achieve quite high accuracy 92.7%–96.2% [29] using freeform alphabetic character recognition, with no text entry evaluation study results. A previous study showed that with the default Xbox 360 gesture based text input interface, the input speed was only 1.83 wpm [30]; with speech recognition for text input, 5.18 wpm text input speed could be achieved [30].

Using a single 3D sensor for gestural text input could be a very challenging task. For example, one typical issue for freehand gesture text input is the difficulty of gesture delimiter design [1], to address this issue, the cross selection technique proposed by Accot and Zhai could be considerd, which enbles selection without clicking action [31]. Studies showed that it could be very promising for freehand selection [32]: users can reach towards the target to select without the to stay and wait for a spesific time. To address the low resolution issue of the 3D sensors, expanding the target could be considered, but most expanding methods don't magnify in motor space, but only in visual space, but not in motor space [33].

### 2.3. Freehand gestural design challenges

Our target is to design freehand gestural interaction to achieve effective and "walk-up-and-use" interaction experiences [1,46,47], while retaining the simplicity and directness of multi-touch interaction, which is very widely used thanks to the smart phones and tablets. Freehand motion tracking enabled by a remote 3D camera is very different from handheld input devices or fiducial markers. The former may involve expensive computer vision tracking task [48,49,50] and real-time accuracy request [51,52]. For example, low cost 3D camera can only track hand motion robustly, but not always accurately track small motions of wrists or fingers from a distance. Besides the challenges of fine movement tracking, there are also some other challenges for freehand interaction, such as noisy body skeleton tracking, no physical button clicking, no surface touching, the absence of physical surface support and no tactile feedback for the hand or fingers.

Based on previous work and our analysis of the characteristics of freehand interaction tracked by low cost 3D cameras, in this paper we take two examples of the most common interaction tasks: selection and text entry to explore the freehand gestural design.

## 3. Study 1: Tile Layout Menu Selection with Freehand Gesture

Menu selection is one of the most basic tasks for interactive systems such as computers, smartphones and TVs. It is very important for users to easily select TV channels, online video services, configurations and other services. Currently a tile layout design is very popular for both mobile devices (e.g. iOS, Android, Windows Phone) and interactive TV/gaming systems (e.g. Apple TV and Xbox). Such a tile layout offers large menu items in a structured interface and supports simple touch interaction.

---

[2]http://www.microsoft.com/en-us/kinectforwindows/

[3]http://www.asus.com/Multimedia/Xtion_PRO/

Used with an interactive TV, a tiled interface and 3D freehand gestures can come together to provide potentially very usable interaction, for example to help users to choose videos, channels or applications. However, very little is known about how best to design such selection interfaces, including features such as target size, selection gesture and the effectiveness of different forms of feedback.

Thus, we conducted an experimental evaluation to investigate some of the main design factors for freehand gestural selection with a tile layout interface. We examined audio and tactile feedback, push gesture depth, tile size and target item position in our experimental design.

### 3.1. Experimental Design

#### 3.1.1. Target Number and Position
In this study we used 9 square tiles in a 3 × 3 layout. In the screen space, the center of the 9 tiles was located at the center of the screen. In motor space, the center (i.e. the position of the hand needed to select the center tile) was directly in front of the user's right shoulder. Hence, the user needed to extend her right hand at shoulder height to select the center target (Fig. 1).

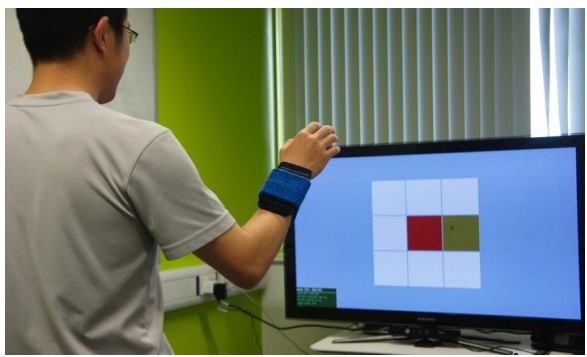

Fig. 1. Experimental Setting.

#### 3.1.2. Feedback
Given the mixed results from previous studies, e.g. [4, 34, 35], it is desirable to achieve better understanding of different forms of feedback with freehand gestural selection and a tile layout. In this study, both tactile and audio feedback were provided using a smart watch mounted on the user's wrist. The tactile feedback used was a 100ms vibration and the audio feedback was a default Android keyboard press sound.

#### 3.1.3. Gesture
There are many possible ways to select a target using freehand gestures. For example, selection can be triggered by dwell time, stroke or reach gestures for different tasks and interfaces [36]. As users are now very familiar with a finger tap gesture on touch screens, we chose a similar movement of pushing forward to select in our study of freehand gestural selection. Hence, we used a push gesture, i.e. moving the hand forward to a specific depth in the Z dimension to select the target.

#### 3.1.4. Depth and Size
A previous study of a freehand virtual keyboard [37] used a push depth of 5cm, and another study suggested that 3cm could be the minimum distance for freehand selection [3]. So in this experiment, we used 4cm as the minimum push depth. As comparisons, we also tested 8cm and 12cm push depth. For target width, we used 12cm, 18cm and 24cm respectively. Depth and width are measured in motor space.

#### 3.1.5. Independent Variables
There were 4 independent variables (distances here are measured in motor space):
- Wearable Feedback (2 levels): Tactile, Audio
- Depth of push gesture (3 levels): 4 cm, 8 cm, 12 cm
- Target Size (3 levels): 12 cm, 18cm, 24 cm length of side
- Target Position (9 levels): Center, Up, Down, Left, Right, Left-Up, Right-Up, Left-Down, Right-Down.

### 3.2. Experiment and Setting

A Samsung PS50C680 50" 3D plasma TV was used with 60 Hz refresh rate at 1280x720 resolution. The height from the ground to the center of the display was 113 cm, and the user stood 200 cm in front of the monitor. The user's movements were tracked using a Microsoft Kinect Version 1 camera, with a refresh rate of 30 fps, and the Kinect for Windows SDK version 1.8 on Windows 8. The Kinect camera was placed at the bottom of the display. The audio and tactile feedback were provided using a Moto G mobile phone running Android 4.4.4, attached to the user's right wrist using a wristband. An Android application was running on the mobile and the feedback was played when the gesture selection was detected. The Android application and Windows application were connected using Wi-Fi (Fig. 1).

*3.3. Participants and Procedure*

We recruited 12 volunteers (8 males and 4 females) ranging from 24 to 37 years. The mean age was 30.25 (sd = 3.96).

Participants were asked to select the target quickly and accurately while remaining relaxed and comfortable. The target was displayed in red and a cursor was rendered as a small black sphere that followed the user's hand motion. The tile currently indicated by the hand was displayed in green. The mapping of movement distance from motor space to screen space was 1.5:1. The order of the target Position was randomized. After the test, the participant answered a questionnaire about her preferences for different Feedback, combination of Depth and Size, and target Position. They rated each measure from 0 (strongly dislike) to 10 (strongly like). Participants were also asked to provide their comments about the different designs. The whole participation took about 30 minutes.

A repeated measures within-participants design was used. There were 6 sessions, 1 for each combination of Feedback and Depth. In each session, there was a practice block of 9 trials (3 trials in each Size with random target Position) and a test block with 81 trials (3 trails for each Position, thus 27 trials for each tile Size). Sessions with the same Feedback were contiguous, and the order of Depth was the same for each Feedback, giving 12 orders of the Feedback and Depth combinations counterbalanced across the 12 participants. The trials with the same tile Size were grouped together in each session and the order of size group was randomized. For each participant, 486 test trials were performed in total.

*3.4. Results*

We recorded hand movement time, error rate and movement distance. A repeated-measures analysis of variance (ANOVA) for Feedback × Depth × Size × Position was used to analyze the movement time, error rate and movement distance, as reported in Fig 2.

*3.4.1. Movement Time*

Main effects were found for Depth ($F_{2, 22}=5.44$, $p<.05$) and Position ($F_{8, 88}=22.13$, $p<.001$). Neither Feedback nor Size had a significant effect ($p>.05$). Interaction effects were found for Depth × Position ($F_{16, 176}=1.94$, $p<.001$) and Size × Position ($F_{16, 176}=3.76$, $p<.001$), as shown in Fig. 2(a).

Post hoc Bonferroni pairwise comparisons showed that the 4 cm push depth was significantly faster than 12 cm.

Down were also good performing positions, significantly faster than Left-Up, Right-Up and Left-Down ($p<.05$).

*3.4.2. Error Rate*

Users were required to select the target successfully, if necessary trying more than once. Each failure to select the correct target was recorded as an error. Main effects were found for Depth[4] ($F_{1.09, 12.00}=27.74$, $p<.001$) and Size[3] ($F_{1.34, 14.78}=13.96$, $p<.01$). Feedback and Position had no significant effect ($p>.05$). An interaction effect was found for Depth × Size ($F_{4, 44}=7.00$, $p<.001$). Mean error rates across Depth and Size are shown in Fig. 2(b).

Post hoc Bonferroni pairwise comparisons showed that Depths of 8 cm and 12 cm were significantly more accurate than 4 cm ($p<.01$), with no significant difference between 8 and 12 cm. On the other hand, Sizes of 12 cm and 18 cm gave significantly more accuracy than 24 cm ($p<.01$), with no significant difference between 12 and 18 cm.

*3.4.3. Movement Distance*

We recorded hand movement distance along the Z dimension per selection. Main effects were found for Depth[3] ($F_{1.29, 14.23}=75.09$, $p<.001$), Size ($F_{2, 22}=13.98$, $p<.001$) and Position ($F_{8, 88}=12.22$, $p<.001$). Feedback had no significant effect. An interaction effect was found for Size × Position ($F_{16, 176}=2.33$, $p<.01$), shown in Fig. 2(c).

Post hoc Bonferroni pairwise comparisons showed that users moved a significantly longer distance with greater Depth ($p<.001$) or larger Size ($p<.05$). Again, the Center position was the best position, where users moved significantly less than for all the other posi-tions (p<.05) except Left and Down. The four corner positions were the worst, significantly longer than the Center and Down positions (p<.05).

---

[4] The sphericity assumption was not met so the Greenhouse-Geisser correction was applied; the corrected degrees of freedom are shown.

We also recorded the overall movement distance in 3D space, and main effects were found for Depth[3] ($F_{1.21, 13.34}$=55.31, p<.001), Size ($F_{2, 22}$=118.14, p<.001) and Position ($F_{8, 88}$=17.03, p<.001). Again, Feedback had no significant effect (p>.05). No interaction effects were found, as shown in Fig. 2(d).

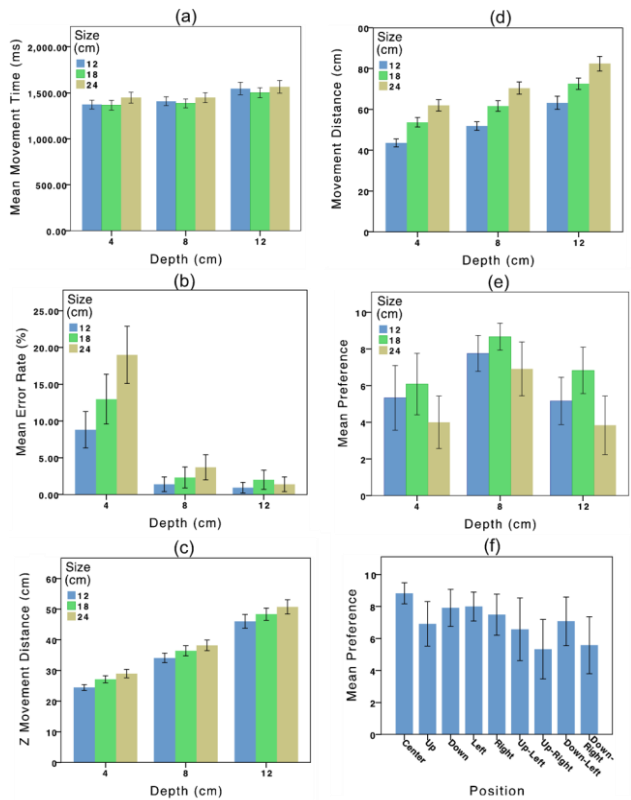

Fig. 2. Results (a) Movement time. (b) Error rate (c) Movement distance in Z combination (d) Overall movement distance (e) Mean preference of Depth and Size (f) Mean preference of the position  (Error bars represent 95% confidence intervals).

### 3.4.4. User Preference

A repeated-measures ANOVA for Depth × Size was used to analyze user preferences. Main effects were found for Depth ($F_{2, 22}$=7.61, p<.01) and Size ($F_{2, 22}$=13.86, p<.001). No interaction effects were found. Post hoc Bonferroni pairwise comparisons showed that users significantly preferred the 8 cm push depth over 4 cm and 12 cm (p<.05), with no difference between 4 cm and 12 cm. For target size, users rated 18 cm significantly higher than the others (p<.05), while 24 cm was significantly less preferred (p<.05), as shown in Fig. 2(e).

A repeated-measures ANOVA for Position was also performed. A main effect was found for Position[3] ($F_{3.53, 38.78}$=4.30, p<.01), while post hoc Bonferroni pairwise comparisons found no significant difference between different positions, as shown in Fig. 2(f).

Analysis using a two-tailed dependent T-test found no preference between the forms of feedback ($t_{11}$=.79, p=.45).

### 3.4.5. Dwell Position

We also recorded the time of user's hand in different depth positions when they perform the selection tasks in different conditions. Results showed that the distribution of the hand position is very similar for different push depth, and most of the time the hands stayed between about 30cm to 90cm. Overall, users' hands stayed in front of their shoulder about 60cm for longest period, as shown in Fig. 3.

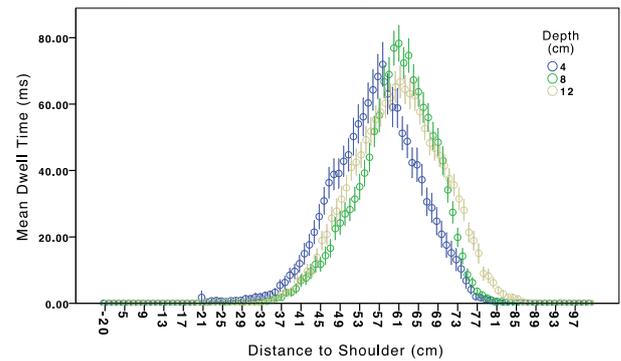

Fig. 3 Dwell position.

### 3.5. Discussion

Consonant with the findings of [4], there was no significant difference in performance or preference between tactile and audio feedback.  Participants' comments, however, identified some qualitative differences between the forms of feedback. For example, several participants really appreciated the privacy of tactile feedback, which could be useful in enabling private interactions with large displays.  The audio feedback, on the other hand, can be shared amongst a group of users, perhaps therefore lending itself more to collaborative interactions with larger displays.

Surprisingly, the bigger target was more error prone than smaller target in the tile layout selected with freehand gestures, especially when the push depth was short, which is very different from the results of previous studies of less densely populated environments [2]. Furthermore, it requires longer hand movement, leading to greater fatigue. We conclude that a large target size in motor space has no benefit for freehand selection with a tiled interface layout and should be avoided.

Although the 4 cm gesture push depth in motor space reduced the selection time a little compared to the 12 cm depth, it introduced many more errors (Fig. 2a and 2b). Thus, a push depth of 4 cm or shorter should be avoided in interaction design. The 8 cm push depth, on the other hand, is fluent and effective for freehand selection according to both quantitative results and user feedback, so it is highly recommended for interaction design. Although a 12 cm push depth shared similar performance with 8 cm, many participants felt that 12 cm was too long and they had to stretch out for it, so a push depth longer than 12 cm is not recommended either. Our user study is based on a 3 × 3 layout. Designers can consider more items if needed, but the item size in motor space should be bigger then 4cm.

Regarding the target position, the middle row at shoulder height showed good performance and user preference. The corner positions were much more difficult to select, especially with a large target, so should be avoided in the interaction design if possible.

**4. Study 2: Freehand Text Entry**

To fully explore the functionality of a smart TV, it is very important for users to have to explore the digital contents by typing the keywords. Speech is a very popular text input method for smart TV at the moment, but speech input can still fail if in some scenarios where the surrounding is noisy the system cannot pick up the voice input very clearly. At the same time, the text can by typed by traditional remote, but this way is usually slow and error prone. Freehand gestural input, on the other hand, could be a potential candidate method for text input for smart interactive TVs. Thus, in this section, we extend our investigation by designing and evaluating freehand text entry methods.

*4.1. Keyboard Layout Design*

Freeform alphabetic character text entry input have been a popular research topic in previous work e.g. [20, 29, 38]. The main challenges for the users using this method is that it could be very difficult to learn and remember a specific set of predefined gestures. For interactive TV user scenarios, it may be too demanding as the users are looking for a quick and easy interaction. Other text entry methods, such as shorthand writing [22, 39], Swype[5] or text input based on speech recognition [30] are based on word prediction, but non-dictionary word is very common for interactive TV scenarios, such as user name, password, email address, or urls. Meanwhile, as virtual keyboard text entry can provide easy learning according to previous studies [19, 21], we mainly investigate virtual keyboard and character based text entry in this section for interactive TVs.

Although character arrangement on a virtual keyboard can be optimized according to the user types and scenarios [19, 24, 26, 40, 41], the Qwerty layout is still the most popular virtual keyboard layout option. It has many benefits [21, 42], and many improved text entry methods are based on the Qwerty layout [42, 43, 44]. Most importantly, it is the most commonly used layout for typing at the moment, which is very important to users when using TVs.

Although freehand can move in 3D space, 3D layout text entry has low performance [21], and 3D selection is less accurate than with a 2D [36], so our design and evaluation are based on the 2D keyboard layout.

*4.1.1. Qwerty Layout Design*

Qwerty is a very common keyboard layout so it is a reasonable option for freehand gestural text input method. For our prototype design and evaluation, we used a Qwerty layout with 28 characters (26 English letters, space and backspace), the design is very similar to the smart phone virtual keyboard (Fig. 4).

*4.1.2. Dual-circle*

We also investigated circle virtual keyboard layout as it may bring benefits for freehand interaction: in circle layout each characters could be accessed easily from the center. For freehand gesture tracked tracked by low cost camera, the character size could be too small in one circle, so for reliable use with noisy tracking input we proposed a Dual-circle layout (Fig 5 and 6). All the characters are evenly distributed within 2 circles. As Qwerty layout is the most common keyboard layout, the character layout is also based on Qwerty: the top and bottom of each circle is used for characters located in the top row and bottom row in Qwerty, and the middle row of the Qwerty layout is turned vertically and put sideways in each circle based on the corresponding hand and fingers when using a Qwerty keyboard. Gaps are used to separate the different character groups for a clearer mapping to the familiar Qwerty layout.

---
[5] http://www.swype.com/

Space and backspace are located in the middle of the keyboard for easy access with both hands. The left-down portion of the right circle and the mirrored portion of the left circle are left blank as users dislike selecting in the left-down direction with the right hand according to previous study [32]. This blank space can be used to hold some less-used special characters if needed.

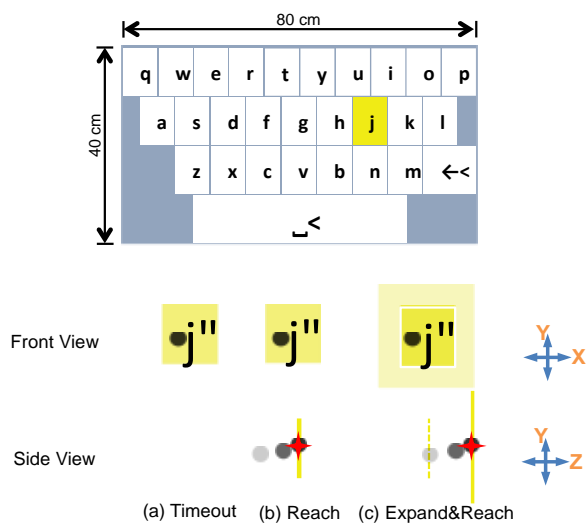

Fig. 4 Up: Qwerty layout. Down: Selection techniques; the spherical grey cursor is controlled by the user's hand position.

### 4.2. Character Selection Design

Freehand gestural selection can be very demanding without the physical button and touch surface. Furthermore, finger and wrist movement are not suitable for freehand interaction enabled by inexpensive cameras when users are standing far away, thus in this section we present some suitable freehand selection techniques for interactive TV.

#### 4.2.1. Timeout

Most commercial freehand selection interface design at the moment is based on the timeout threshold. Such design is easy to be understood and performed by uses when they just start to use freehand selection. However, the selection dwell time can slow down the selection performance. For both keyboard layouts design in our paper, the timeout technique can be used: the user points to a character by the X-Y position of her hand and waits for the timeout threshold, e.g. 1.2 s, to select the character (Fig. 4.a, Fig. 5.a).

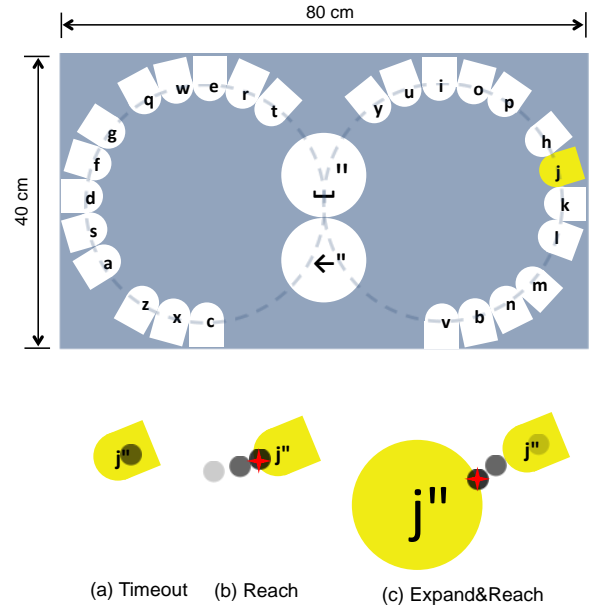

Fig. 5 Up: Dual-circle layout. Down: Selection techniques; spherical grey cursor controlled by the user's hand position.

#### 4.2.2. Reach

Users can select by moving their hands to reach into the target in 3D [32]. For the 2 different virtual Qwerty keyboard layout, the reach selection is designed as below:

- QWERTY: the user can point to the character (X-Y movement), then reach forward (Z-dimension movement) to select (Fig. 4.b).
- Dual-circle: the user can move the hand (X-Y movement) to reach across the border of the character (Fig. 5.b).

#### 4.2.3. Expand&Reach

We also designed Expand&Reach techniques for both keyboard layouts for easier selection and error tolerance.

- Qwerty: when the user points to a character, an expanded target appears along the Z-dimension (e.g. 5 cm further away than the current hand position and 2 times bigger than original size). The user moves her hand forward to reach the expanded target in order to select (Fig. 4.c).
- Dual-circle: when the user points to a character, an expanded character tab containing

the target character appears in the center of the corresponding circle (Fig. 6). The user moves her hand to reach the expanded target to select it (Fig. 5.c).

*4.3. Experimental Evaluation*

*4.3.1. Independent variables*
- Layout (Qwerty, Dual-circle)
- Selection Techniques (Timeout, Reach, Expand&Reach)
- Day (1 to 5).

*4.3.2. Participants*
6 volunteers (4 males, 2 females) were recruited from the local campus. Their mean age was 26 (sd = 1.7), and they are all right handed with some experience of gestural interaction, mainly for gaming.

*4.3.3. Procedure*
The evaluation lasted for 5 days with 6 sessions every day. Each session tested a combination of Layouts and Selection techniques. In each session, 4 sentences were presented for the participant to reproduce, the first as practice followed by 3 test sentences. Six sets of sentences were randomly selected from MacKenzie and Soukoreff's phrase sets [45] and were assigned randomly to different sessions. User preferences and NASA task load index (TLX) data were collected on the first and last days.

Each character is white and highlights yellow when pointed at. With Timeout selection, the color gradually changes from yellow to green until timeout. Using Reach and Expand&Reach with Qwerty, the character gradually changes from yellow to green as the hand moves forward to reach it. The selected character appears immediately below the target sentence. A "typing" sound is played when a character is entered correctly. If the entry is incorrect, an error sound is played instead and the input is shown in red. All mistakes must be corrected for each sentence.

Both keyboards were 80 cm x 40 cm, with the top edge at the same height as the user's shoulders in motor space. Two spheres sized 2 cm were controlled with the hands. With the Qwerty layout, the hand in front of the other is enabled and rendered in black, the other is rendered grey and disabled to avoid accidental selection. With the Dual-circle layout, the blank center area is large enough to accommodate an idle hand without accidental selection, so no disable mechanism was used. For timeout selection technique with both layouts, the dwell time was 1.2 s. The 1.2 s dwell time was based on a pilot study which showed that less than 1.2 s produced more errors, and the observation that almost all commercial Kinect interfaces with timeout selection use more than 1.5 s.

*4.3.4. Experimental Setting*
A Sanyo PDG-DWL2500 3D projector was used at 1280 x 720 resolution with a 203 x 115 cm screen centered at 130 cm height to simulate a large interactive TV display. A Microsoft Kinect camera was used with a refresh rate of 30 fps and the Kinect for Windows SDK V1.5 on Windows 7. The Kinect camera was placed 50 cm in front of the screen at a height of 70 cm. The user stood 250 cm from the screen (Fig. 6).

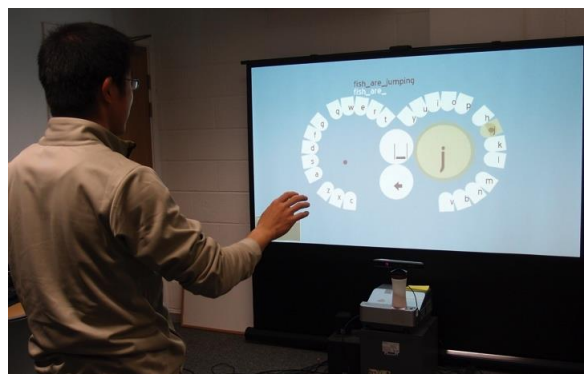

Fig. 6. Experimental setting.

*4.4. Results*

*4.4.1. Typing Speed*
We use repeated-measures ANOVA for Layout x Selection Technique x Day to analyze the text input speed. We found main effects were for Selection Technique ($F_{2,10}$=144.27, p<.001) and Day ($F_{4,20}$=21.49, p<.001). No significant effect was found for layout ($F_{1,5}$=1.38, p=.29). Interaction effects were found for Layout x Selection Technique ($F_{2,10}$=11.69, p<.01) and Day x Selection Technique ($F_{8,40}$=6.05, p<.001).

Post hoc Bonferroni pairwise comparisons were also used to analyze the the text input speed. Results showed that Reach and Expand&Reach were both significantly faster than Timeout (p<.001). Text input speed in the last 3 days was significantly faster than on the first day (p<.05). Fig. 7 and Table 1 show mean text input speeds.

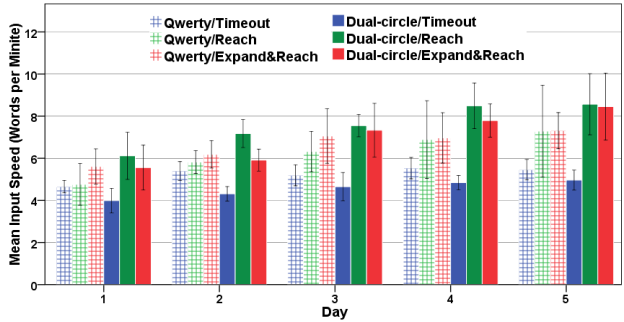

Fig. 7. Mean text input speed. In this and later charts, error bars represent 95% confidence intervals.

Table 1. Mean speed (wpm) in day 1, day 5 and over 5 days.

|  | Qwerty | | | Dual-circle | | |
|---|---|---|---|---|---|---|
|  | Time-out | Reach | Expand &Reach | Time-out | Reach | Expand &Reach |
| Day 1 | 4.65 | 4.76 | 5.61 | 3.99 | 6.11 | 5.56 |
| Day 5 | 5.46 | 7.29 | 7.31 | 4.96 | 8.57 | 8.46 |
| 5 days overall | 5.25 | 6.22 | 6.63 | 4.55 | 7.58 | 7.01 |

*4.4.2. Error Rate*

We used a repeated-measures ANOVA for Layout x Selection Technique x Day to analyze error rate. We found main effects for Layout ($F_{1,5}$=10.86, p<.05) and Selection Technique ($F_{2,10}$=11.09, p<.01). No significant effect ($F_{4,20}$=1.94, p=.14) was found for Days. And there was no interaction effect. Fig. 8 and Table 2 show mean error rates.

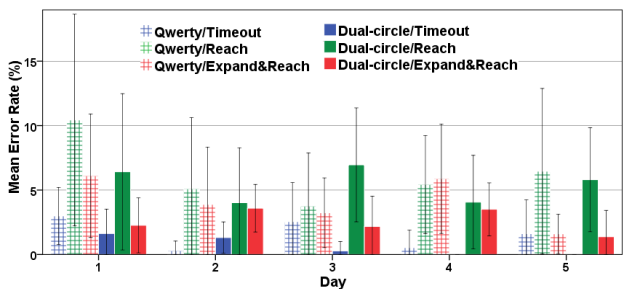

Fig. 8. Mean error rate.

We also used post hoc Bonferroni pairwise comparisons to analyze the error rate. We found significantly more error-prone in QWERTY than Dual-Circle (p<.05). We also found that timeout had significantly fewer errors than Reach (p<.05). No other significant differences were found between the selection techniques.

Table 2. Mean error rate (%) in day 1, day 5 and over 5 days.

|  | Qwerty | | | Dual-circle | | |
|---|---|---|---|---|---|---|
|  | Time-out | Reach | Expand &Reach | Time-out | Reach | Expand &Reach |
| Day 1 | 3.00 | 10.44 | 6.11 | 1.63 | 6.42 | 2.27 |
| Day 5 | 1.61 | 6.46 | 1.60 | 0.00 | 5.81 | 1.38 |
| 5 days overall | 1.60 | 6.24 | 4.14 | 0.64 | 5.45 | 2.58 |

*4.4.3. Hand Movement in 3D Space*

The hand movement distance per character was also recorded in day 5. We used a repeated-measures ANOVA for Layout x Selection Technique to analyze hand movement distance. We found main effects for Layout ($F_{1,5}$=39.42, p<.01) and Selection Technique ($F_{2,10}$=10.21, p<.01). And there was no interaction effect.

We also used post hoc Bonferroni pairwise comparisons to analyze the hand movement distance. Results showed that Qwerty layout had significantly more hand movement distance than Dual-Circle (p<.01). The Timeout selection technique had significantly less movement distance than Reach (p<.05). No other significant differences between selection techniques were found. Fig. 9(a) shows mean hand movement distance per character.

One-way ANOVA for six text entry methods is used to analyze the hand movement distance in different axes (i.e. X, Y, Z) using. No significant effect of axis (p<.05) was found for Qwerty/Timeout and Dual-circle/Reach. Main effects were found for Axis (p<.001) with Qwerty/Reach, Qwerty/Expand&Reach and Dual-circle/Timeout.

With Qwerty/Reach and Qwerty/Expand&Reach methods, post hoc Bonferroni pairwise comparisons showed hand movement in the Z axis was significantly more than in the X and Y axes (p<.05), on the other hand, with Dual-circle/Timeout and Dual-circle/Expand&Reach methods, movement in the Z axis was significantly less (p<.01). Fig. 9 (b) shows the mean hand movement distance.

*4.4.4. Task Load*

We used a repeated-measures ANOVA for Layout x Selection Technique x Day to analyze the NASA task load index (TLX). We found main effects Layout ($F_{1,5}$=9.21, p<.05), Selection Technique ($F_{2,10}$=5.75, p<.05) and Day ($F_{1,5}$=7.61, p<.05). We found no interaction effects.

We also used post hoc Bonferroni pairwise comparisons to analyze the TLX index. The results showed that the TLX index of Qwerty layout was significantly higher than the Dual-circle layout (p<.05), the TLX index of reach selection technique was significantly higher than Expand&Reach (p<.05), and the TLX index on day 1 was significantly higher than on day 5 (p<.05). The overall workload results were shown in Fig 10.

circle/Expand&Reach in both days as the error tolerance design allowed the users to be more relax when selecting, reducing the physical effort and mental attention.

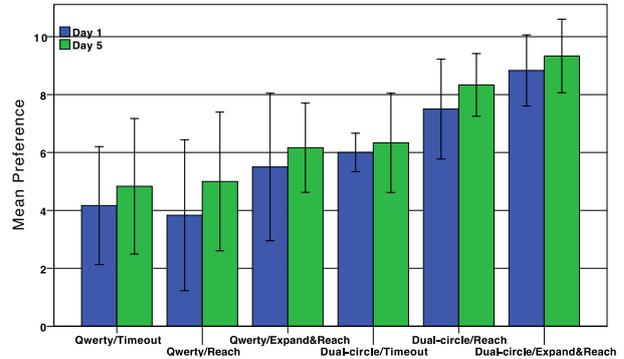

Fig. 11. User preference

### 4.5. Discussion

Overall, Dual-circle/Expand&Reach was the best text input method. Thanks to the Expand&Reach selection technique, the users make fewer errors. It also leaves enough blank space in the center to help the users to relax their hands. So it can be a practical freehand gestural text entry method. The Dual-circle/Reach method, on the other hand, was found to be error-prone. With the improvement tracking resolution of the remote motion sensors, it could be a fast entry method. Although users are more familiar with the Qwerty keyboard, the dual-circle has better performance and lower task load. This is mainly because that with QWERTY, the characters are packed in 2D, so the selection is more difficult. While with the dual-circle layout, they can just relax the hand because there is enough blank space in the center to prevent careless error selection.

When using the Qwerty/Reach text entry method, users reported high physical demand. With Qwerty/Expand&Reach, on the other hand, as relative location was used, it was easier to select. However, both Qwerty/Reach and Qwerty/Expand&Reach required hand movement for character selection along the Z dimension, leading to long movement distance when inputting text (Fig. 9). In contrast, with the dual-circle layout, there no need to move hand along the Z dimension so it felt easier to use.

With Reach and Expand&Reach techniques, users noticed their improvements with both keyboard layouts. Especially with the Dual-circle layout, typing speed increased continuously over 5 days (Fig. 7). With timeout selection, on the other hand, the dwell

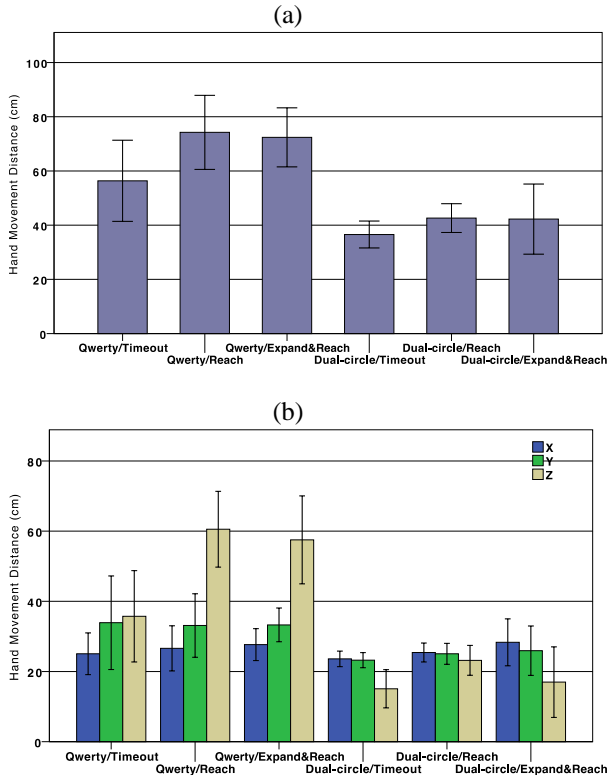

Fig. 9. Mean hands movement distance per character. (a) Hand movement distance in 3D space (b) Hand movement distance in X, Y and Z axis.

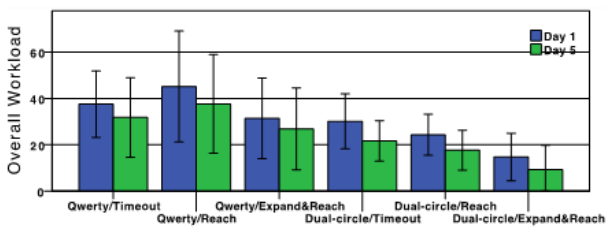

Fig. 10. NASA task load index (TLX) results.

#### 4.4.5. User Preference

We collected the preference on the first and fifth days. All participants preferred Dual-time slowed text entry speed so there is almost no performance improvement.

## 5. Conclusion

In this paper, we investigated the design of freehand gestural interaction for interactive TV based on 2 common user tasks of menu selection and text input. We looked at many design factors including tactile/audio feedback, gesture depth, menu item size and position, as well as virtual keyboard layout and key selection methods. Our studies suggest a set of interesting findings and design guidelines for freehand gestural interaction.

*5.1. Comparisons to Previous Work*

We found both similarities and differences compared to previous work. For example, the user performance and preference between tactile and audio feedback showed no significant difference, which is similar to [4]. Users didn't like to move their hands forward and backward frequently, which is similar with the findings in [54], as such movement requires larger muscle groups (e.g. the shoulder) thereby making the actions slow and tiring.

More interestingly, we also found some major differences from previous work. One of the main differences is that bigger menu items may create more errors when using a freehand push gesture to select. This is largely due to the users' need to move their hand across other menu items before reaching the target. When the menu items are larger, the hands need to move a greater distance before reaching the target, and have more chances to make errors. As their hands are moving freely in the air with no support or guidance from physical surfaces (as is provided by the desktop for mouse input and by the tablet screen for touch input), the users may more easily trigger the accidental selection of other menu items before they reach the target.

Another interesting difference is that in our study, the results suggested the performance of QWERTY layout was worse than circle layout, on the other hand, previous work using hand held devices for text input showed the opposite [21]. Such differences are duo to different input devices: moving the bare hand and arm freely in the air is very different from existing input methods. The characteristics of freehand gestural interaction, such as the lack of physical support for the hand and the absence of button clicking, introduce very different user task performance and require very careful considerations from designers, especially if they try to leverage previous designs for desktop input devices as bases for the design of freehand gestural interaction with an interactive TV.

*5.2. Other Design Suggestions*

From the findings of our studies, we also synthesized some practical design suggestions for freehand interaction with interactive TVs:

1. Target size should not be too large in motor space (in our case, should not be larger than 18 cm for a 3 x 3 menu). Large target size is not recommended as it will lead not just to longer hand movement, but will also introduce more errors. This is also very different from previous design guidelines for 2D interaction devices, indicating that freehand gestural interface design has its own unique characteristics.

2. If using a push gesture for selection, the push depth should be about 8 cm for a 3 x 3 menu design. This is the setting that gave the best combination of performance and user preference. A long push depth is not recommended due to the fatigue issue; and a short push depth may introduce more errors.

3. User preference suggests that frequently used items should be positioned in the interface at shoulder height, and less frequently used items positioned in other rows if possible.

4. Frequent movements in the Z dimension are not recommended. This is because a) the hand moves more slowly forward and backward [53]; b) actions in the Z dimension can be error-prone [36]; and c) frequent movements in the Z dimension can increase physical demands.

5. The circle layout could be considered more with freehand gestural interaction design. This is because in the circle layout, all items can be reached directly without moving the hands over other objects. And the large blank area in the center can allow users to relax their hands without worrying about triggering undesired actions, thus potentially reducing both arm fatigue and errors.

6. Combination of expanding target and reach is useful for densely populated selection tasks. The target can expand not only in virtual space but also in motor space, so it is well suited to freehand interaction techniques due to its error tolerance.

Freehand gesture is a promising interaction technique for interactive TV. This paper reports our investigations of the design and evaluation of such freehand gestural interaction. Our evaluations reveal some unique characteristics of freehand gestural interaction. And in our experiment the hand movements are all higher than user's waist. So we can expect similar results if users are sitting.

In our experiment, users are standing when performing gestures. As most freehand gesture SDKs (e.g. Microsoft Kinect SDK) can support sit-down tracking mode quite well, and the hands movements in our studies are all above waist, we can expect similar performance when users are sitting down. Future work includes investigation of other design factors, such as the amplitude of tactile and audio feedback, different virtual keyboard layout and supporting different interaction tasks such as manipulating 3D virtual objects.

## 6. Acknowledgement

The author W. Li is supported by the UK EPSRC project EP/K023578/1.